\documentclass[10pt,a4paper]{article}
\usepackage{jcappub}
\usepackage{pdflscape}
\usepackage{amsmath}
\usepackage{amssymb}
\usepackage{dcolumn}
\usepackage{bm}
\usepackage{color}
\usepackage{epsfig}
\usepackage{amsfonts}
\usepackage{graphicx}
\usepackage{subfigure}
\usepackage{dcolumn}
\begin{document}

%%%%%%%%%%%%%%%%%%%%%%%%%%%%%%%%%%%%%%%%%%%%%%%%%%%%%%%%
\title{Using Dominant and Weak Energy Conditions for build New Classe of Regular Black Holes}
%%%%%%%%%%%%%%%%%%%%%%%%%%%%%%%%%%%%%%%%%%%%%%%%%%%%%%%%

\author[a,b]{Manuel E. Rodrigues}  
\author[a,c]{Ednaldo L. B. Junior}
\author[a]{Marcos V. de S. Silva}

\affiliation[a]{ Faculdade de F\'{\i}sica, PPGF, Universidade Federal do Par\'{a}, 66075-110, Bel\'{e}m, Par\'{a}, Brazil}

\affiliation[b]{Faculdade de Ci\^{e}ncias Exatas e Tecnologia, Universidade Federal do Par\'{a}\\
Campus Universit\'{a}rio de Abaetetuba, CEP 68440-000, Abaetetuba, Par\'{a}, Brazil}

\affiliation[c]{Faculdade de Engenharia da Computa\c{c}\~{a}o, 
Universidade Federal do Par\'{a}, Campus Universit\'{a}rio de Tucuru\'{\i}, 
68464-000, Tucuru\'{\i}, Par\'{a}, Brazil}

\emailAdd{esialg@gmail.com}
\emailAdd{ednaldobarrosjr@gmail.com}
\emailAdd{marco2s303@gmail.com}

%%%%%%%%%%%%%%%%%%%%%%%

\abstract{We study the generalization of the mass function of classes of regular spherically symmetric black holes solutions of in $ 4D $ coming from the coupling of General Relativity with Non-linear Electrodynamics, through the requirement of the energy conditions. Imposing that the solution must be regular and that the Weak and Dominant Energy Conditions are simultaneously satisfied, for models with the symmetry $T^0_{\;\;0} = T^1_{\;\;1} $ of the tensor energy-momentum, we construct a new general class of regular black holes of which two cases are asymptotically Reissner-Nordstrom.}

\maketitle

%%%%%%%%%%%%%%%%%%%%%%%

\section{Introduction}
\label{sec1}
%%%%%%%%%%%%%%%%%%%%%%%%%%

Since the beginning of General Relativity solutions of black holes has gained a lot of attention. A black hole has the following basic characteristics \cite{wald}: It is a solution of Einstein's equations and there is a surface that separates the causal structure of space-time, called the event horizon. These solutions initially always presented a singularity covered by this horizon, thus representing a great difficulty of physical interpretation. In 1935, Einstein and Rosen suggested a small modification \cite{visser}, changing the signal of the contribution of the electromagnetic field, to formulate a geometry that could escape the singularity within the event horizon, thus creating the wormholes. This persistence in obtaining a solution in which geodesics were not to be interrupted by a singularity continued until a suggestion of Bardeen \cite{bardeen}, known as Bardeen's regular black hole. This new structure was always regular in all space-time, not presenting singularity in the curvature scalars and their contractions, as $R$, $R_{\mu\nu}R^{\mu\nu}$ and $\mathcal{K}=R_{\mu\nu\alpha\beta}R^{\mu\nu\alpha\beta}$, besides having an event horizon. 
\par 
Many solutions of regular black holes have arisen since then, which always present some characteristics in common, as it is the case of Bardeen. Most solutions have spherical symmetry \cite{rbh1}-\cite{JAR} and there are some with axial symmetry \cite{rbh2}-\cite{Mustapha}. We can also have regular solutions in modified gravity theory
 \cite{manuel}-\cite{manuel3}. 
\par 
There are some key features in getting new solutions for regular black holes. One is that the metric and the curvature scalars do not have divergence in all space-time. This in spherical symmetry is sufficient to ensure regularity \cite{wald}. Another feature that arises when trying to avoid the appearance of a singularity, the energy distribution is similar to that of a de Sitter-like solution, with $p_r=p_t=-\rho$ (Sakarov criterion \cite{sakarov}) for $p$ being the radial or tangential pressure, and $\rho$ the energy density, clearly violating the Strong Energy Condition (SEC), $\rho+3p\geq 0$. Thus, a good possibility to obtain new solutions appears in the coupling of the gravitation with the call Non-Linear Electrodynamics (NED) \cite{peres}, where this equation of state is common. The Bardeen's regular black hole can be interpreted as a solution coming from an NED \cite{beato}. There are also regular solutions that behave asymptotically as Reissner-Nordstrom or others that do not have this characteristic, as is the case of Bardeen. Another important characteristic is that coming from the so-called energy conditions. For solutions with spherical symmetry, it can be shown that every regular solution of black hole violates in some region the SEC \cite{zaslavskii}, but the Dominant Energy Condition (DEC) and Weak Energy Condition (WEC) can always be met, as is the case of Bardeen which always satisfies WEC.
\par 
We have new possibilities for attempting to construct solutions of regular black holes with spherical symmetry that always satisfy a WEC, this has been started in the work to Balart et al \cite{balart}. The idea is that one can generalize quite to obtain regular solutions through the requirement to always satisfy WEC. Following this direction started in \cite{balart}, we can show that there are still several possibilities for new solutions of regular black holes, which can be asymptotically Reissner-Nordstrom or not. From the requirement to satisfy WEC, regularity and to be asymptotically Reissner-Nordstrom, one can construct metrics much more general than those already presented in the literature. This is the central idea in this work.
\par 
The structure of the work is divided as follows. In section \ref{sec2} we present the structure of the Einstein's equations related to the models of regular solutions and the WEC, DEC and SEC. In section \ref{sec3} we obtain a new class of solutions of regular black holes through the exigences of the previous section.  We present our conclusions and perspectives on section \ref{sec4}.

%%%%%%%%%%%%%
\section{The energy conditions and the regular black hole solutions}
\label{sec2}
%%%%%%%%%%%%%%%%%

We will briefly present how the possible models of regular black holes are related to agreement with WEC and DEC.
\par 
Let us first consider a metric that is spherically symmetrical and static
\begin{eqnarray}
ds^2=e^{a(r)}dt^2-e^{b(r)}dr^2-r^2\left(d\theta^2+\sin^2\theta d\phi^2\right)\label{ele}\,
\end{eqnarray}
where $a(r)$ and $b(r)$ are arbitrary functions of the radial coordinate. The action of the theory is given by the coupling between gravitation and NED
\begin{eqnarray}
S=\int d^4x\sqrt{-g}\left[R+16\pi G \mathcal{L}_{NED}\right]\label{action}\,,
\end{eqnarray}
where $R$ is the curvature scalar and $\mathcal{L}_{NED}\equiv \mathcal{L}_{NED}(F)$ is an arbitrary function of the scalar $F=(1/4)F_{\mu\nu}F^{\mu\nu}$, with $F_{\mu\nu}=\partial_{\mu}A_{\nu}-\partial_{\nu}A_{\mu}$ being the Maxwell tensor. This theory falls to Maxwell for the specific choice
 $\mathcal{L}_{NED}=F$.  One known form of formulation for regular black holes is to introduce a mass function $M(r)$ through component $g_{11}$ of the metric by
\begin{eqnarray}
e^{-b(r)}=1-\frac{2M(r)}{r}\label{M}\,.
\end{eqnarray}
Making the functional variation of the action \eqref{action} in relation to the metric we have the following Einstein's equations
\begin{eqnarray}
T^{0}_{0}=T^{1}_{1}=\frac{1}{4\pi G r^2}\frac{dM}{dr}\,,\,T^{2}_{2}=T^{3}_{3}=\frac{1}{8\pi G r}\frac{d^2M}{dr^2}\,,\label{EMT}
\end{eqnarray}
where
\begin{eqnarray}
&&T_{\mu\nu}=g_{\mu\nu}\mathcal{L}_{NED}-\frac{\partial\mathcal{L}_{NED}(F)}{\partial F}F_{\mu}^{\;\;\beta}F_{\nu\beta}\,.
\end{eqnarray}
We have used $a(r)=-b(r)$ which arises from the spherical symmetry and the equation $G^{0}_{0}-G^{1}_{1}=0$ (symmetry of the energy-momentum tensor $T^0_0=T^1_1$). We can identify $T^{0}_{0}=\rho$ and $T^{1}_{1}=-p_r,T^{2}_{2}=T^{3}_{3}=-p_t$, with $\rho$ being the energy density and $p_r$ and $p_t$ are the radial and tangential pressure respectively. Now we can establish the energy conditions \cite{visser}
\begin{eqnarray}
&&SEC(r)=\rho+p_{r}+2p_{t}\geq 0\,,\\
&&WEC_{1,2}(r)=\rho+p_{r,t}\geq 0\;,\\
&& WEC_{3}(r)=\rho\geq 0, \\ &&
DEC_{2,3}(r)=\rho-p_{r,t}\geq 0\;,
\end{eqnarray}
where, in view of the identities $DEC_1(r)\equiv WEC_3(r)$ and $NEC_{1,2}(r)\equiv WEC_{1,2}(r)$, this conditions are not written. Performing coordinate transformation $r=1/x$, from \eqref{EMT} we have
\begin{eqnarray}
\rho(x)=-\frac{x^4}{4\pi G}\frac{dM}{dx}=-p_r, p_t=-\frac{x^4}{8\pi G}\left(2\frac{dM}{dx}+x\frac{d^2M}{dx^2}\right)\label{eq1}\,.
\end{eqnarray}
Then we see directly $WEC_{1}(x)=0$ which is always satisfied. We also have $DEC_2(x)=2WEC_3(x)$. So let's focus on the energy conditions
\begin{eqnarray}
&&SEC(x)=x^4\left(2\frac{dM}{dx}+x\frac{d^2M}{dx^2}\right)\leq 0\,,\label{sec}\\
&&WEC_{2}(x)=x^4\left(4\frac{dM}{dx}+x\frac{d^2M}{dx^2}\right)\leq 0\;,\label{wec2}\\
&& WEC_{3}(x)=x^4\frac{dM}{dx}\leq 0, \label{wec3}\\ &&
DEC_{3}(x)=x^5\frac{d^2M}{dx^2}\geq 0\;.\label{dec}
\end{eqnarray}

Now we will focus only on solutions that always satisfy the WEC and DEC. If we require the WEC and the regularity of the metric we must have \cite{balart}
\begin{eqnarray}
x\rightarrow +\infty , -x^4\frac{dM}{dx}\rightarrow c_1\,,\label{cond1}
\end{eqnarray}
with $c_1$ being a positive constant. We still have if the solution behaves asymptotically as Reissner-Nordstrom \cite{balart}
\begin{eqnarray}
x\rightarrow 0, -\frac{dM}{dx}\neq 0 \,.\label{condRN}
\end{eqnarray}
In the next section we'll show how to build new classes of regular black hole solutions.
 
%%%%%%%%%%%%%%%%%%%%%%%%%%%%
\section{The new class of regular black hole solutions}\label{sec3}
%%%%%%%%%%%%%%%%%%%

The work of Balart et al suggests a mass function $ M (x) $, which behaves asymptotically as Reissner-Nordstrom and is general as
\begin{eqnarray}
-\frac{dM}{dx}=\frac{c_1}{(1+c_2x^{\alpha})^{4/\alpha}}\,,\label{balart}
\end{eqnarray}
with $c_1$ and $c_2$ being positive constants, and $\alpha$ positive integer. Actually $\alpha$ need not be positive integer if we do not require the solution to be asymptotically Reissner-Nordstrom. We can verify that the $WEC_{1,2,3}(x)$ and the $DEC_{1,2,3}(x)$ are always satisfied for the model
 \eqref{balart} and $SEC(x)$ is violated to some region, as expected \cite{zaslavskii}. 
\par 
We can now show that there are several other possibilities of generalization of this mass function. Taking the requirements \eqref{cond1}, \eqref{condRN} and the WEC We can construct the following mass function
\begin{eqnarray}
-\frac{dM}{dx}=\frac{c_1(c_8+c_3x^{c_4}+c_5x^{c_6})}{(1+c_2x^{\alpha})^{(4+c_7)/\alpha}}\,,
\end{eqnarray}
that integrating from infinity to $x$, for $Re[\alpha]\geq 0$, $Re[c_4-c_7]<3$ and $Re[c_6-c_7]<3$, results in
\begin{eqnarray}
&&M(x)=c_1c_2^{-(4+c_7)/\alpha}x^{-3-c_7}\Big\{\frac{c_8}{3+c_7}\;_{2}F_{1}\Big[d_1,d_2,d_3,-\frac{x^{-\alpha}}{c_2}\Big]+\frac{c_3x^{c_4}}{3-c_4+c_7}\; _{2}F_{1}\Big[d_2,d_4,d_6,-\frac{x^{-\alpha}}{c_2}\Big]\nonumber\\
&&+\frac{c_5x^{c_6}}{3-c_6+c_7}\; _{2}F_{1}\left[d_2,d_5,d_7,-\frac{x^{-\alpha}}{c_2}\right]\Big\}\,,\label{sol1}\\
&&d_1=\frac{3+c_7}{\alpha},d_2=\frac{4+c_7}{\alpha},d_3\frac{3+c_7+\alpha}{\alpha},d_{4,5}=\frac{3-c_{4,6}+c_7}{\alpha},d_{6,7}=\frac{3-c_{4,6}+c_7+\alpha}{\alpha}.
\end{eqnarray}
with $_2F_1[i,k,l,z]$ being the Gauss's Hypergeometric function. The mass function \eqref{sol1} presents the following characteristics: does not introduces divergences in the metric; may compose a metric that is asymptotically Reissner-Nordstrom; and generates solutions that satisfy the energy conditions NEC, DEC e WEC. The mass function also can presents a important restriction in the asymptotic limit. We will explain this better. Astrophysical black holes are usually modeled through axially symmetrical and charged solutions, like Kerr-Newman. These black holes present a physical restriction $q^2<<m^2$ \cite{Punsly}, where $m=\lim\limits_{x\rightarrow 0}M(x)$ is the $ADM$ mass for asymptotically flat solutions. So that, we may consider the mass function \eqref{sol1} as a toy model for an astrophysical  solution with spherical symmetry (without rotation), where the $ADM$ mass must have the following restrictions $10^{-17}M_{\odot}<m<10^{2}M_{\odot}$ for primordial black holes \cite{Bellomo}, and $3M_{\odot}<m<10^{2}M_{\odot}$ for detections of gravitational waves from the merger of binary black holes by the LIGO Collaboration \cite{Cholis}. Actually we can even more restrict the mass for $3M_{\odot}<m<40M_{\odot}$ \cite{Fishbach}. 
\par 
The mass function \eqref{sol1} recover some regular black holes solutions already known. For $c_3=c_5=c_7=0$, $c_8=1$, $c_1=q^2/2$, $c_2=(\pi q^2/(8m))^2$ and $\alpha=2$ we have the regular black hole of I. Dymnikova \cite{Irina}. With $c_5=0$, $c_4=c_3=c_7=1$, $c_8=0$, $c_1=3mg^2$, $c_2=g^2$, $\alpha=2$ we obtained the Bardeen regular black hole. For $c_5=0$, $c_3=1$, $c_4=c_7=2$, $c_8=0$, $c_1=6(ml)^2$, $c_2=2ml^2$, $\alpha=3$ we have the regular solution of S. A. Hayward \cite{Hayward}. We construct a more general solution than \eqref{sol1} in the appendix A. For the rest of this work we will make $c_8=1$.
\par 
The Lagrangian $\mathcal{L}_{NED}$, its derivative $\mathcal{L}_{F}\equiv \partial\mathcal{L}_{NED}/\partial F$ and the electric field $F^{10}$, for the mass function \eqref{sol1}, are given, in terms of the radial coordinate, by\footnote{To find $\mathcal{L}_{NED}$ and $\mathcal{L}_F$ we solve the Einstein's equations in terms of these functions. To determine the electric field $F^{10}$, we solve the modified Maxwell equations $\nabla_{\mu}\left(F^{\mu\nu}\mathcal{L}_F\right)=0$, with the electric charge $q$ as an integration constant.} 
\begin{eqnarray}
&&\mathcal{L}_{NED}=\frac{c_1 r^{c_7+\alpha} }{\kappa ^2\left(c_2+r^{\alpha}\right)^{\frac{c_7+4+\alpha}{\alpha }}}  \Bigg(\frac{c_3}{r^{c_4}} \left((c_4+c_7+2)
\frac{c_2}{r^{\alpha }}-c_4-2\right)+\nonumber\\
&&+\frac{c_2}{r^\alpha} \left( (c_7-c_6+2)
\frac{c_5}{r^{c_6}}+c_7+2\right)- (c_6+2)\frac{c_5}{r^{c_6}}-2\Bigg),\\
&&\mathcal{L}_F=\frac{\kappa ^2 q^2 \left(\frac{1}{r}\right)^{c_7+4+\alpha} 
	\left(c_2+r^{\alpha }\right)^{\frac{c_7+4+\alpha}{\alpha }}}{c_1 \left( \frac{c_3}{r^{c_4}}\left(\frac{c_2}{r^{\alpha }} (c_4-c_7) +c_4+4\right)+ \frac{c_5}{r^{c_6}} \left(
	\frac{c_2}{r^{\alpha}}(c_6-c_7) +c_6+4\right)- \frac{c_2 c_7}{r^{\alpha }}+4\right)},\\
&&F^{10}=\frac{c_1 r^{c_7+2+\alpha} }{\kappa ^2 q\left(c_2+r^{\alpha}\right)^{\frac{c_7+4+\alpha}{\alpha }}}\Bigg( \frac{c_3}{r^{c_4}} \left((c_4-c_7)
	\frac{c_2}{r^{\alpha }}+c_4+4\right)+\nonumber\\
	&&+\frac{c_5}{r^{c_6}} \left( (c_6-c_7)\frac{c_2}{r^{\alpha}}+c_6+4\right)- \frac{c_2 c_7}{r^{\alpha }}+4\Bigg),
\end{eqnarray}
with $\kappa^2=8\pi G$.

\par 
Taking the mass function \eqref{sol1} the WEC, DEC and SEC on \eqref{wec2}-\eqref{dec} are
\begin{eqnarray}
&&WEC_2(x)=-c_1x^{4}(1+c_2x^{\alpha})^{-d_2}(1+c_3x^{c_4}+c_5x^{x_6})\\
&&WEC_3(x)=-c_1x^{-c_7-\alpha}\Big[4-c_2c_7x^{\alpha}+c_3x^{c_4}(4+c_4+c_2x^{\alpha}(c_4-c_7))\nonumber\\
&&+c_5x^{c_6}(4+c_6+c_2x^{\alpha}(c_6-c_7))\Big],\\
&&SEC(x)=-c_1x^{4}(1+c_2x^{\alpha})^{-(d_2+1)}\Big[2-c_2x^{\alpha}(2+c_7)+c_3x^{c_4}(2+c_4+c_2x^{\alpha}(c_4-2-c_7))+\nonumber\\
&&+c_5x^{c_6}(2+c_6+c_2x^{\alpha}(c_6-2-c_7))\Big],\\
&&DEC_3(x)=c_1x^4(1+c_2x^{\alpha})^{-(d_2+1)}\Big[c_2x^{\alpha}(4+c_7)-c_3x^{c_4}(c_4+c_2x^{\alpha}(c_4-4-c_7))+\nonumber\\
&&-c_5x^{c_6}(c_6+c_2x^{\alpha}(c_6-4-c_7))\Big].
\end{eqnarray}
We see that the SEC can never be satisfied in all space-time, always in some region it will be violated, being consistent with the result of \cite{zaslavskii}. 
\par 
If we look for solutions that only satisfy the WEC and DEC we will find solutions that can be singular, since the general mass function \eqref{sol1} is not restricted to only regular solutions. Therefore, we must first ensure regularity in all space-time. For this, we will calculate the curvature scalar and Kretschmann for the following metric
\begin{eqnarray}
&&ds^2=\left(1-2xM(x)\right)dt^2-x^{-4}\left(1-2xM(x)\right)^{-1}dx^2-x^{-2}\left[d\theta^2+\sin^2\theta d\phi^2\right]\,,
\end{eqnarray}
which results in
\begin{eqnarray}
&&R(x)=-2x^5\frac{d^2M}{dx^2}\label{R}\,,\\
&&\mathcal{K}=R_{ijkl}R^{ijkl}=4x^{10}\left(\frac{d^2M}{dx^2}\right)^2+\nonumber\\
&&+32x^9\frac{dM}{dx}\frac{d^2M}{dx^2}+16x^8M\frac{d^2M}{dx^2}+80x^8\left(\frac{dM}{dx}\right)^2+96x^7M\frac{dM}{dx}+48x^6M^2.\label{Kre}
\end{eqnarray}

If there is a divergence in the curvature scalar then there will also be a divergence in the Kretschmann scalar. For spherically symmetrical solutions it is then convenient to first check the regularity of the curvature  scalar. Thus, by taking the mass function \eqref{sol1}, the curvature scalar becomes
\begin{eqnarray}
&&R(x)=\frac{2c_1c_2^{-d_2}}{(1+c_2x^{\alpha})}x^{-c_7}\left(1+\frac{x^{-\alpha}}{c_2}\right)^{-d_2}\Big[-c_2(4+c_7)x^{\alpha}+c_3x^{c_4}(c_4+c_2(c_4-(4+c_7))x^{\alpha})\nonumber\\
&&+c_5x^{c_6}(c_6+c_2(c_6-(4+c_7))x^{\alpha})\Big]\,.
\end{eqnarray}

We can see that in the asymptotic limit $x\rightarrow 0$ we have $R(x)\sim x^4\rightarrow 0$, for $c_{4,6}\geq 0$. Now for the solution to be regular in origin $x\rightarrow +\infty$, we have to check the power of $x$ that govern. Then in this limit we have the following powers that govern in the numerator of the curvature scalar $\alpha+c_4-c_7$ and $\alpha+c_6-c_7$. Considering $c_4\geq c_6$, then the highest power is  $\alpha+c_4-c_7$, which subtracting from the denominator, which is $\alpha$, become $c_4-c_7$. For these considerations the dominant power of the curvature scalar, for this limit, becomes  $R(x)\sim x^{c_4-c_7}$. Then the curvature scalar goes to zero or to a constant for $c_4\leq c_7$. So let's do $c_7=c_4+k$, with $k\geq 0$, this ensures that the solution is regular at $x\rightarrow +\infty$. 
\par 
Let's now check the energy conditions for $c_4\geq c_6$ and   $c_7=c_4+k$, with $k\geq 0$
\begin{eqnarray}
&&WEC_2(x)=-c_1c_2^{-d_2}x^{-c_4-k}\left(1+\frac{x^{\alpha}}{c_2}\right)^{-d_2}(1+c_3x^{c_4}+c_5x^{c_6}),\label{eq1}\\
&&WEC_3(x)=-\frac{c_1c_2^{d_2}}{(1+c_2x^{\alpha})}x^{-c_4-k}\left(1+\frac{x^{\alpha}}{c_2}\right)^{-d_2}[4-c_2(c_4+k)x^{\alpha}+c_3x^{c_4}(4+c_4-c_2kx^{\alpha})+\nonumber\\
&&+c_5x^{c_6}(4+c_6+c_2(c_6-(c_4+k))x^{\alpha})]\\
&&DEC_3(x)=\frac{c_1c_2^{-d_2}}{(1+c_2x^{\alpha})}x^{-c_4-k}\left(1+\frac{x^{\alpha}}{c_2}\right)^{-d_2}[c_2(4+c_4+k)x^{\alpha}-c_3x^{c_4}(c_4-c_2(4+k)x^{\alpha})+\nonumber\\
&&-c_5x^{c_6}(c_6+c_2(c_6-(4+c_4+k))x^{\alpha})].\label{eq2}
\end{eqnarray}
We can see that the $WEC_2(x)$ is always satisfied. However, in the $WEC_3(x)$, the term $[-c_2c_3kx^{\alpha+c_4}]$ can never be compensated, when $k>0$, for it be satisfied. Then we take
 $k=0$, which implies $c_7=c_4$. Now is the term $[-c_2c_4x^{\alpha}]$ which must be compensated, which is solved for the choice $c_7=c_4=\alpha$. So the WEC and DEC can be fulfilled to $(4+\alpha)c_3\geq \alpha c_2$ and $c_6\geq \alpha$, but previously we imposed $c_4=\alpha\geq c_6$, which implies $c_6=c_4=c_7=\alpha$, what is not interesting. Then we have the only possibility more appreciable in that $c_5=0$ the WEC and DEC are satisfied for $[\alpha/(4+\alpha)]c_3\leq c_2\leq [(4+\alpha)/\alpha]c_3$. Then we take $c_2=[(4+\alpha)/\alpha]c_3$, and the WEC and DEC are always fulfilled. The mass function of this solution is given by
\begin{eqnarray}
&&M(x)=c_1\left(\frac{c_3(4+\alpha)}{\alpha}\right)^{-(4+\alpha)/\alpha}\Big\{\frac{c_3}{3}x^{\alpha}\;_{2}F_{1}\left[\frac{3}{\alpha},\frac{4+\alpha}{\alpha},\frac{3+\alpha}{\alpha},-\frac{\alpha x^{-\alpha}}{c_3(4+\alpha)}\right]+\nonumber\\
&&+\frac{1}{3+\alpha}\;_{2}F_{1}\left[\frac{3+\alpha}{\alpha},\frac{4+\alpha}{\alpha},2+\frac{3}{\alpha},-\frac{\alpha x^{-\alpha}}{c_3(4+\alpha)}\right]\Big\}.
\end{eqnarray} 
Let's check the properties for the particular case where $\alpha=1$. The curvature scalar, the Kretschmann scalar and the mass function are given by
\begin{eqnarray}
&&M(x)=\frac{c_1(4+5c_3x)}{75c_3(1+5c_3x)^4}\,,\label{M1}\\
&&R(x)=-\frac{8c_1c_3x^5(6+5c_3x)}{(1+5c_3x)^6}\,,\\
&&\mathcal{K}(x)=\frac{32c_1^2x^6}{1875c_3^2}\left(\frac{8-120c_3x+2325c_3^2x^2+22500c_3^4x^4+37500c_3^5x^5+15625c_3^6x^6}{(1+5c_3x)^{12}}\right)\,.
\end{eqnarray}
We then have the following limits 
\begin{eqnarray}
&&\lim_{x\rightarrow 0}\{M(x),R(x),\mathcal{K}(x)\}=\left\{\frac{4c_1}{75c_3},0,0\right\},\\
&&\lim_{x\rightarrow +\infty}\{M(x),R(x),\mathcal{K}(x)\}=\left\{0,-\frac{8c_1}{3125c_3^4},\frac{32c_1^2}{29296875c_3^8}\right\}.
\end{eqnarray}
We see here that this is a regular solution in all space-time and asymptotically flat. We can expand the mass function on Taylor series to $x<<1$
\begin{eqnarray}
&&1-2xM(x)\cong1-\frac{8c_1}{75c_3}x+2c_1x^2+O(x^3)=1-2mx+q^2x^2+O(x^3)\,,
\end{eqnarray}
with $m$ being the ADM mass and $q$ the electric charge. When $c_1=q^2/2$ and $c_3=[2q^2/(75m)]$ the solution becomes asymptotically Reissner-Nordstrom. We can verify the same behavior for all cases of integer $\alpha$. We still have the possibility of fractional $\alpha$.
\par 
The functions $\mathcal{L}_{NED}$, $\mathcal{L}_F$ and $F^{10}$, in terms of the radial coordinate, are written as
\begin{eqnarray}
\mathcal{L}_{NED}(r)&=&\frac{10125m^4q^2\left(4q^4+180mq^2r-1125m^2r^2\right)}{\kappa^2\left(15mr+2q^2\right)^6},\\
\mathcal{L}_F(r)&=&\frac{\kappa^2\left(15mr+2q^2\right)^6}{22781250m^6r^6},\\
F^{10}(r)&=&\frac{22781250qm^6r^4}{\kappa^2\left(15mr+2q^2\right)^6}.
\end{eqnarray}
The Lagrangian and the electric field are regular in all space-time. In the appendix B we will use this solution to show a way to construct regular solutions in the $f(R)$ theory.
\par 
Now we can turn our attention to another direction to construct solutions that satisfy WEC and DEC. For the condition of regular solution in which $c_7=c_4$ and $c_4\geq c_6$ in \eqref{eq1}-\eqref{eq2}, with $k=0$, we can see that the term $[-c_2c_4x^{\alpha}]$ can be compensated through $c_6=\alpha$. This generates a new term, $[c_2c_4c_5x^{2\alpha}]$, which must be compensated for WEC and DEC to be met. So we can choose $c_4=c_7=2\alpha$ and thus the WEC and DEC are satisfied to $[\alpha/(2(2+\alpha))]c_5\leq c_2\leq [(4+\alpha)/(2\alpha)]c_5$ and $[\alpha/(2(2+\alpha))]c_2c_5\leq c_3\leq [(4+\alpha)/(2\alpha)]c_2c_5$. In short, when $c_4=c_7=2\alpha,c_6=\alpha$, $[\alpha/(2(2+\alpha))]c_5\leq c_2\leq [(4+\alpha)/(2\alpha)]c_5$ and $[\alpha/(2(2+\alpha))]c_2c_5\leq c_3\leq [(4+\alpha)/(2\alpha)]c_2c_5$, the solution satisfies WEC and DEC. The mass function becomes
\begin{eqnarray}
&&M(x)=c_1c_2^{-2(2+\alpha)/\alpha}x^{-3-2\alpha}\Big\{\frac{1}{3+2\alpha}\;\;_2F_1\left[2+\frac{3}{\alpha},2+\frac{4}{\alpha},3+\frac{3}{\alpha},-\frac{x^{-\alpha}}{c_2}\right]+\nonumber\\
&&+\frac{c_3}{3}x^{2\alpha}\;_2F_1\left[2+\frac{4}{\alpha},\frac{3}{\alpha},\frac{3+\alpha}{\alpha},-\frac{x^{-\alpha}}{c_2}\right]+\frac{c_5x^{\alpha}}{3+\alpha}\;_2F_1\left[2+\frac{4}{\alpha},\frac{3+\alpha}{\alpha},2+\frac{3}{\alpha},-\frac{x^{-\alpha}}{c_2}\right]\Big\}.
\end{eqnarray} 
\par 
Let's look at a particular case for $\alpha=1,c_2=2c_5$ and $c_3=5c_5^2$. In this case the mass function, curvature scalar and Kretschmann are
\begin{eqnarray}
&&M(x)=\frac{c_1(4+10c_5x+25c_5^2x^2)}{30c_5(1+2c_5x)^5}\,,\label{M2}\\
&&R(x)=-\frac{2c_1c_5x^5(11+40c_5^2x^2)}{(1+2c_5x)^7},\\
&&\mathcal{K}(x)=\frac{8c_1^2x^6}{75c_5^2(1+2c_5x)^{14}}(8-16c_5x+492c_2^2x^2-314c_5^3x^3+\nonumber\\
&&+6313c_5^4x^4-2170c_5^5x^5+22650c_5^6x^6+10000c_5^8x^8).
\end{eqnarray}
We then have the following limits 
\begin{eqnarray}
&&\lim_{x\rightarrow 0}\{M(x),R(x),\mathcal{K}(x)\}=\left\{\frac{2c_1}{15c_5},0,0\right\},\lim_{x\rightarrow +\infty}\{M(x),R(x),\mathcal{K}(x)\}=\left\{0,-\frac{5c_1}{8c_5^4},\frac{25c_1^2}{384c_3^8}\right\}.
\end{eqnarray}
We see here again that this is a regular solution in all space-time and asymptotically flat. We can expand the mass function on Taylor series to $x<<1$
\begin{eqnarray}
&&1-2xM(x)\cong1-\frac{4c_1}{15c_5}x+2c_1x^2+O(x^3)=1-2mx+q^2x^2+O(x^3)\,.
\end{eqnarray}
When $c_1=q^2/2$ and $c_5=[q^2/(15m)]$ the solution is asymptotically Reissner-Nordstrom. We can verify the same behavior for all cases of integer $\alpha$. We still have the possibility of fractional $\alpha$.
\par 
The functions $\mathcal{L}_{NED}$, $\mathcal{L}_F$ and $F^{10}$, for the radial coordinate, becomes
\begin{eqnarray}
\mathcal{L}_{NED}(r)&=&\frac{253125m^4q^2\left(4q^6-42mq^4r+225m^2q^2r^2-1350m^3r^3\right)}{2\left(2q^2+15mr\right)^7\kappa^2},\\
\mathcal{L}_F(r)&=&\frac{2\left(2q^2+15mr\right)^7\kappa^2}{759375m^5r^5\left(28q^2+15mq^2r+900m^2r^2\right)},\\
F^{10}(r)&=&\frac{759375m^5qr^3\left(28q^2+15mq^2r+900m^2r^2\right)}{2\left(2q^2+15mr\right)^7\kappa^2}.
\end{eqnarray}
The Lagrangian and the electric field are well behaved in all space-time.
\par 
Turning our attention to the WEC and DEC \eqref{eq1}-\eqref{eq2}, we see that another possibility arises when $c_7=c_4=\alpha$ and $c_6=-\alpha$. In this case the WEC and DEC are fulfilled to $[\alpha/(4+\alpha)]c_3\leq c_2\leq [(4+\alpha)/\alpha]c_3,4\geq\alpha$ and $4\geq 2c_2c_5\alpha$. The mass function becomes
\begin{eqnarray}
&&M(x)=c_1c_2^{-(4+\alpha)/\alpha}x^{-3-\alpha}\Big\{\frac{c_5x^{-\alpha}}{3+2\alpha}\;_2F_1\left[2+\frac{3}{\alpha},\frac{4+\alpha}{\alpha},3+\frac{3}{\alpha},-\frac{x^{-\alpha}}{c_2}\right]+\nonumber\\
&&+\frac{c_3}{3}x^{\alpha}\;_2F_1\left[\frac{3}{\alpha},\frac{4+\alpha}{\alpha},\frac{3+\alpha}{\alpha},-\frac{x^{-\alpha}}{c_2}\right]+\frac{1}{3+\alpha}\;_2F_1\left[\frac{3+\alpha}{\alpha},\frac{4+\alpha}{\alpha},2+\frac{3}{\alpha},-\frac{x^{-\alpha}}{c_2}\right]\Big\}.
\end{eqnarray}
Taking the example $c_2=[(4+\alpha)/\alpha]c_3,\alpha=2$ we have
\begin{equation}
M(x)=\frac{c_1}{72} \left(\frac{45 c_3 x^4 (27 c_3 c_5-2)+3 x^2 (225 c_3 c_5-14)+72 c_5}{x \left(3 c_3 x^2+1\right)^2}+\frac{5
	\sqrt{3} (2-27 c_3 c_5) \cot ^{-1}\left(\sqrt{3c_3} x\right)}{\sqrt{c_3}}\right),
\end{equation}
and the limits 
\begin{eqnarray}
&&\lim_{x\rightarrow 0}\{M(x),R(x),\mathcal{K}(x)\}=\left\{\frac{c_1}{c_3^3}sing[c_3]^3sign[c_5]\infty,0,0\right\},\\
&&\lim_{x\rightarrow +\infty}\{M(x),R(x),\mathcal{K}(x)\}=\left\{0,-\frac{8c_1}{27c_3^2},\frac{32c_1^2}{2187c_3^4}\right\}.
\end{eqnarray}
We see then that the solution is asymptotically non-flat, but regular in all space-time. There are still solutions with
 fractional $\alpha$  which are asymptotically flat.

The functions $\mathcal{L}_{NED}$, $\mathcal{L}_F$ and $F^{10}$, for the coordinate $r$, are given by 
\begin{eqnarray}
&&\mathcal{L}_{NED}=\frac{2 c_1 \left(3 c_3^2+r^4 (9 c_3 c_5-1)+4 c_3r^2\right)}{\kappa ^2 \left(3 c_3+r^2\right)^4},\\
&&\mathcal{L}_F=\frac{\kappa ^2 q^2 \left(3 c_3+r^2\right)^4}{2 c_1 r^8 \left(-6mc_3 c_5+c_5 r^2+2\right)},\\
&&F^{10}=\frac{2 c_1 r^6 \left(-6 c_3 c_5+c_5 r^2+2\right)}{\kappa ^2 q \left(3 c_3+r^2\right)^4}.
\end{eqnarray}
The Lagrangian and the electric field are regular in all space-time.
%%%%%%%%%%%
\section{Conclusion}\label{sec4}
%%%%%%%%%%%%%%

We assumed that the derivative of the mass function derived from the coupling of General Relativity with Non-linear Electrodynamics, with symmetry $T^0_{\;\;0}=T^1_{\;\;1}$, it should be finite for the asymptotic limit of radial infinity and be proportional to a positive power (negative for $x=1/r$) in the limit of the origin of the radial coordinate. This led us to construct a more general mass function than that of the Balart case, which for some limit of the parameters may fall in this particular case.
\par 
Imposing regularity in all space-time we arrive at a relation between the constants of the mass function, which imposing the WEC and DEC, results in three solutions that have two free constants, which were related to the mass ADM and the electric charge, for asymptotically flat cases. Two classes  of solutions are asymptotically flat and asymptotically Reissner-Nordstrom, in the case $c_6=-\alpha$ the solution is asymptotically non-flat only for  $\alpha$ integer.
\par 
Solutions logically violate SEC in some region. But we believe that there are still other classes of regular solutions that simultaneously satisfy WEC and DEC, coming from the mass function \eqref{sol1}.

%%%%%%%%%%%
\vspace{1cm}

%%%%%%%%%%%%%%
{\bf Acknowledgements}: M. E. R.  thanks Conselho Nacional de Desenvolvimento Cient\'ifico e Tecnol\'ogico - CNPq, Brazil, Edital MCTI/CNPQ/Universal 14/2014  for partial financial support.

%%%%%%%%%%%%%%%%%%%%%%

%%%%%%%%%%%%%%%%%%%%%%%%%
\section*{Appendix A: More general mass function}\label{apA}
%%%%%%%%%%%%%%%%%%%%%%%%

We may consider a more general model of mass function where
\begin{eqnarray}
-\frac{dM(x)}{dx}=c_1\frac{\left(c_8+\Sigma_{i=1}^{N}k_{i}x^{n_i}\right)}{(1+c_2x^{\alpha})^{(4+c_7)/\alpha}}\,,
\end{eqnarray}
that integrating results in
\begin{eqnarray}
&&M(x)=c_1c_2^{-(4+c_7)/\alpha}x^{-3-c_7}\Bigg\{\frac{c_8}{3+c_7}  \;_{2}F_{1}\left[\frac{3+c_7}{\alpha},\frac{4+c_7}{\alpha},\frac{3+c_7+\alpha}{\alpha},-\frac{x^{\alpha}}{c_2}\right]+\nonumber\\
&&+\Sigma^{N}_{i=1}\frac{k_ix^{n_i}}{3+c_7-n_i}\;_{2}F_{1}\left[\frac{4+c_7}{\alpha},\frac{3+c_7-n_i}{\alpha},\frac{3+c_7+\alpha-n_i}{\alpha},-\frac{x^{\alpha}}{c_2}\right]\Bigg\}\,.
\end{eqnarray}

The constants in the mass function must be adjusted to satisfy the energy conditions and the restriction $3M_{\odot}<m<40M_{\odot}$ to be valid ($m=\lim\limits_{x\rightarrow 0}M(x)$).

%%%%%%%%%%%%%%%%%%%%%%%%%
\section*{Appendix B: Generalization to $f(R)$ Gravity}
%%%%%%%%%%%%%%%%%%%%%%%%
In this appendix, we will show a way to find regular solutions in $f(R)$ theory coupled to the Non-linear Electrodynamics. The action that describes this theory is given by:
\begin{equation}
S_{f(R)}=\int d^4x\sqrt{-g}\left[f(R)+16\pi G \mathcal{L}_{NED}\right].
\end{equation}
The equations of motion may be written as
\begin{eqnarray}
&&\hspace{-.3cm}R_{\mu\nu}-\frac{1}{2}g_{\mu\nu}R=
f_R^{-1}\big[8\pi G T_ { \mu\nu }
+\frac{1}{2}g_{\mu\nu}\left(f-Rf_R\right) -\left(g_{\mu\nu}\square-\nabla_{
	\mu}\nabla_{\nu}\right)f_R\big] \equiv 8\pi G
  \mathcal{T}_{\mu\nu}^{(eff)}\label{energyeff},
\end{eqnarray}
where $f_R\equiv d f(R)/d R$. To construct the solutions we consider the line element \eqref{ele} with the ansatz \eqref{M}. From the components \eqref{energyeff} and assume $a=-b$, we can write \cite{manuel2}
\begin{equation}
f_R(r)=c_1r+c_0,
\label{fR}
\end{equation}
that recover General Relativity for $c_0=1$ and $c_1=0$. The function $f(R)$ is obtained, in terms of the radial coordinate, through
\begin{equation}
f(r)=\int f_R(r)\frac{dR(r)}{dr}dr,
\label{fr}
\end{equation}
where the curvature scalar in given in terms of the mass function as
\begin{eqnarray}
R(r)&=&-\frac{2 r M''+4 M'}{r^2}.
\end{eqnarray}

From \eqref{energyeff} and the modified Maxwell equations, the functions $\mathcal{L}_{NED}$, $\mathcal{L}_F$ and $F^{10}$ are
\begin{eqnarray}
&&\mathcal{L}_{NED}(r)=-\frac{1}{16\pi Gr^2}\left[r^2f(r)+4c_0M'+2c_1r\right],\label{LNL}\\
&&\mathcal{L}_F(r)=-\frac{8q^2\pi G}{r^2}\left[\left(c_1r+c0\right)rM''-\left(2c_0+c_1r\right)M'-3c_1M+c_1r\right]^{-1},\label{LFNL}\\
&&F^{10}(r)=\frac{1}{8q\pi G}\left\{3c_1M+\left(2c_0+c_1r\right)M'-r\left[c_1+\left(c_0+c_1r\right)M''\right]\right\}.\label{F10fr}
\end{eqnarray}

In this sense, all the relevant expression are written in terms of the mass function. So that, for which different mass function we have a different $f(R)$ theory and a different Non-linear Electrodynamics. Is also important verify if the solutions satisfy the energy conditions. However, from the theorem proved in \cite{manuel3}, the energy conditions are identical from the General Relativity.

As an example, we will use the mass function \eqref{M1} that we construct in the section \ref{sec3} for $\alpha=1$ with $c_1=q^2/2$ and $c_3=[2q^2/(75m)]$.

\begin{figure}
	\begin{center}
		\includegraphics[scale=0.6]{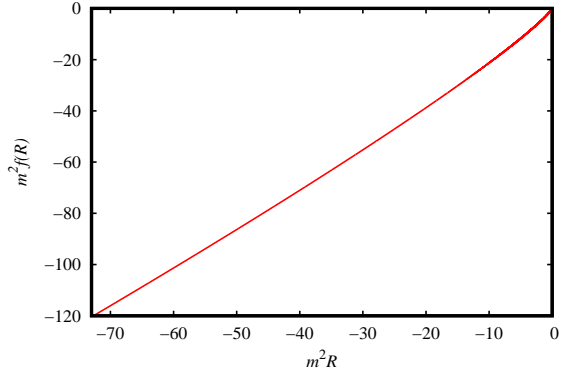}
		\vspace{0.5cm}
		\caption{Parametric plot for $f(R)$ as a function of $R$, with $q=0.1m$, $c_1=2m^{-1}$ e $c_0=1$.}
		\label{fs2}
	\end{center}
\end{figure}

Due the symmetry $a=-b$, the curvature scalar is given by
\begin{eqnarray}
R(r)=-\frac{162000 m^4 \left(45 m q^4 r+q^6\right)}{\left(15 m r+2 q^2\right)^6}.
\end{eqnarray}
That is the same from General Relativity, so that does not present divergences.
From the equation \eqref{fr}, we obtain
\begin{equation}
f(R)=-\frac{2700 m^3 q^4 \left[60 c_0 m \left(45 m r+q^2\right)+c_1 \left(3375 m^2 r^2+180 m q^2 r+4 q^4\right)\right]}{\left(15 m r+2 q^2\right)^6}.
\end{equation}
In fig. \ref{fs2} we show a parametric plot for $f(R)\times R$. In this we can see the non-linear behavior of this theory.

The functions $\mathcal{L}_{NED}$, $\mathcal{L}_F$ and $F^{10}$ becomes
\begin{eqnarray}
&&F^{10}(r)=\frac{r}{\kappa ^2 q \left(15 m r+2 q^2\right)^6} \bigg\{22781250 c_0 m^6 q^2 r^3\hspace{-0.09cm}-\hspace{-0.09cm}c_1 \hspace{-0.09cm}\left[r^4m^5\hspace{-0.09cm}\left(11390625 m r\right.\hspace{-0.1cm}+\hspace{-0.1cm}9112500 q^2\right)\hspace{-0.1cm}(r\hspace{-0.1cm}-\hspace{-0.1cm}3m) \nonumber\\
&&+3037500 m^4 q^4 r^4+540000 m^3 q^6 r^3\left.+54000 m^2 q^8 r^2+2880 m q^{10}  r+64 q^{12}\right]\bigg\},\label{f10FR}\\
&&\mathcal{L}_{NED}(r)=-\frac{1}{\kappa ^2 r \left(15 m r+2 q^2\right)^6}\bigg\{10125c_0 m^4 q^2 r \left(1125 m^2 r^2-180 m q^2 r-4 q^4\right)\nonumber\\
&&+c_1 \big[64 q^{12}+2880 m q^{10} r-5400 m^2 q^8 r (m-10 r) -27000 m^3 q^6 r^2 (9 m-20 r)\nonumber\\
&&-1518750 m^4 q^4 r^3(3 m-2 r)+9112500 m^5 q^2 r^5+11390625 m^6 r^6\big]\bigg\},\\
&&\mathcal{L}_{F}(r)=-\frac{\kappa ^2 q^2 \left(15 m r+2 q^2\right)^6}{r^3} \bigg\{c_1 \big[64 q^{12}+2880 mq^{10} r+54000 q^6 r^2m^2(10 m  r+ q^2)\nonumber\\
&&+3037500 m^4 q^4 r^4-(11390625 m r+9112500 q^2)(3 m-r)m^5r^4\big]-22781250c_0 m^6 q^2 r^3\bigg\}.
\end{eqnarray}
If $c_0=1$ and $c_1=0$ we recover the result of General Relativity.
The method present in this section can be used to construct much more regular solutions. Some examples are described in \cite{manuel2,manuel3}.

%%%%%%%%%%%%%

\end{document}